\documentclass[preprint,12pt,nofootinbib]{revtex4}
\usepackage[paper=letterpaper,margin=1.5in]{geometry}
\usepackage{epsfig,color}
\usepackage{graphicx}
\usepackage{amssymb,amsmath}
\usepackage{time}
\usepackage[varg]{txfonts}
\usepackage{hyperref}
\hypersetup{
	colorlinks,
	citecolor=black,
	linkcolor=black}

\renewcommand{\vec}[1]{\textnormal{\boldmath$#1$}}
\begin{document}
	
\begin{titlepage}
		\begin{flushright}
			{\Large 
			DESY-18-038\\
			March 2018}
		\end{flushright}
		\vspace{3cm}
		\centering
		{\huge\bfseries Analytical Impedance Models\\ 
			for	Very Short Bunches\par}
		\vspace{2cm}
		{\scshape\Large Igor Zagorodnov\par}
		{\Large\itshape Deutsches Elektronen-Synchrotron, Hamburg, Germany \par}
		\vfill
		{ Invited talk\\ 
			at ICFA Mini-Workshop on Impedances\\ 
			and Beam Instabilities in Particle Accelerators\\ 
			18 - 22 September 2017\\ 
			Benevento, Italy\\}
		\vspace{1cm}
		submitted for publication in CERN Yellow Reports
\end{titlepage}

\mbox{}
\thispagestyle{empty}
\newpage

\title{Analytical Impedance Models for	Very Short Bunches}

\author{Igor Zagorodnov\footnote{Igor.Zagorodnov@desy.de}}

\affiliation{Deutsches Elektronen-Synchrotron, Hamburg, Germany}

\date{\today}

\begin{abstract}
	We discuss several analytical models for impedances of very short bunches. The approximate analytical models are compared with direct solution of Maxwell's equations.
\end{abstract}

\maketitle

\section{Introduction}

We consider only relativistic case where the longitudinal $w_{\parallel}$ and the transverse $\vec w_{\perp}$ wake functions for a relativistic point charge $q$ are defined as~\cite{Chao93}
\begin{equation}\label{EqWakeDef}
w_{\parallel}(s)=-\frac {1}{q}\int_{-\infty}^{\infty}E_z(z,t)|_{t=\frac {z+s}{c}}dz,\quad
\frac{\partial}{\partial s}\vec w_{\perp}= \nabla w_{\parallel}.
\end{equation}
The coupling impedance is given by the Fourier transform of the wake function
\begin{align}
\vec Z(k)=\frac {1}{c} \int_0^{\infty} \vec w(s) e^{i k s}ds,\nonumber
\end{align}
where $c$ is a velocity of light.	

The difficulty of numerical calculation of wakefields can be assotiated with a small parameter $\sigma_z/a$, where $\sigma_{z}$ is the rms bunch length and $a$  is the typical size of the structure. Indeed, for a closed structure of typical size $a$ the calculation time of the wake potential with finite-difference code~\cite{ECHO} is proportional to $(a/\sigma_z)^4$. 

If the structure is open, f.e. supplied with an outgoing pipe, then the calculation time increases considerably as we have to propagate the field in the outgoing pipe along the formation length  $a^2/\sigma_z$ to reach an accurate estimation of the improper integral in Eq.(\ref{EqWakeDef}). For the typical parameters of the European FEL linac~\cite{EXFEL}, the rms bunch length $\sigma_z=25$µm and the aperture radius $a=35$mm, the formation length of the wake potential (transient region) is approx. 25m. Application of an “indirect integration” method~\cite{Zag06} allows to replace the improper integral in Eq.(\ref{EqWakeDef}) with a proper one in the outgoing pipe cross-section. It returns the calculation time back to $(a/\sigma_z)^4$, but the numerical burden remains huge for very short bunches, $\sigma_z<<a$.

On the other hand the small parameter $\sigma_z/a$ allows to develop  asymptotic analytical models and to avoid time-consuming numerical simulations. At this paper we will review several analytical models for the impedances of very short relativistic bunches and compare them with direct numerical solution of Maxwell's equations.

\section{Optical Model}

In order to estimate the high frequency impedance of short transitions  an optical model was developed in~\cite{Opt07_1,Opt07_2}. In this approximation we assume that the electromagnetic fields carried by a short bunch propogate along straight lines equivalent to rays in the geometric optics. An obstacle inside the beam pipe can intercept the rays and reflect them away from their original direction. The energy in the reflected rays is assotiated with the energy radiated by the beam, which can then be related to the impedance. 
\begin{figure}[!htb]
	\centering
	\includegraphics*[width=300pt]{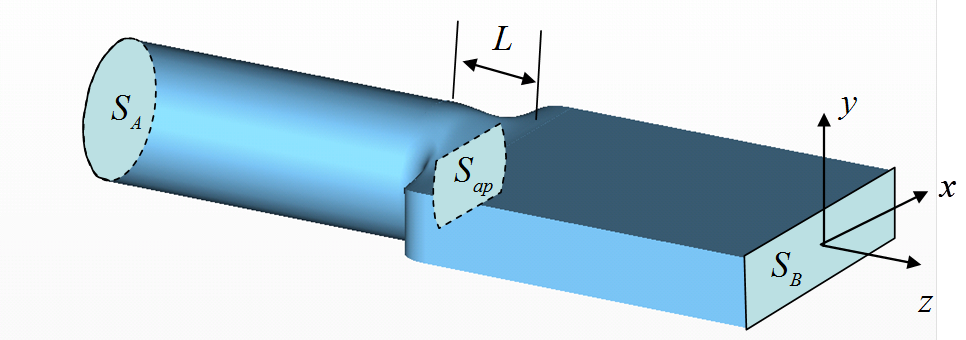}
	\caption{An example of transition geometry.}
	\label{Fig01}
\end{figure}

Consider a short transition with aperture $S_{ap}$  between two pipes with apertures $S_A$   and $S_B$   as shown in Fig.~\ref{Fig01}. Let $a$  is a characteristic size of the aperture $S_{ap}$. If the bunch has a short rms length $\sigma_z$, $\sigma_z<<a$, and the transition length $L$ between the ingoing pipe aperture $S_A$ and the outgoing pipe aperture $S_B$ is much shorter than the formation length, $L<<a^2/\sigma_z$, then  the high frequency longitudinal  impedance is a constant which can be calculated by relation
\begin{equation}\label{EqOptImp}
Z_{\parallel}(\vec r_1,\vec r_2)=-\frac {2 \epsilon_0}{c}  \int_{\partial S_{ap}} \phi_B (\vec r_2,\vec r)\partial_{\vec n} \phi_A (\vec r_1,\vec r)dl,
\end{equation}
where $\vec n$ is the outward pointing unit normal to the line element $dl$, $\epsilon_0$ is  the permittivity of free space, $\vec r_1$ and $\vec r_2$ are  offsets of the leading and the trailing particles, correspondingly, and  $\phi_A, \phi_B$   are the Green’s functions for the Laplacian in the ingoing and the outgoing pipe cross-sections (see ~\cite{Opt07_1} for details).
\begin{figure}[!htb]
	\centering
	\includegraphics*[width=300pt]{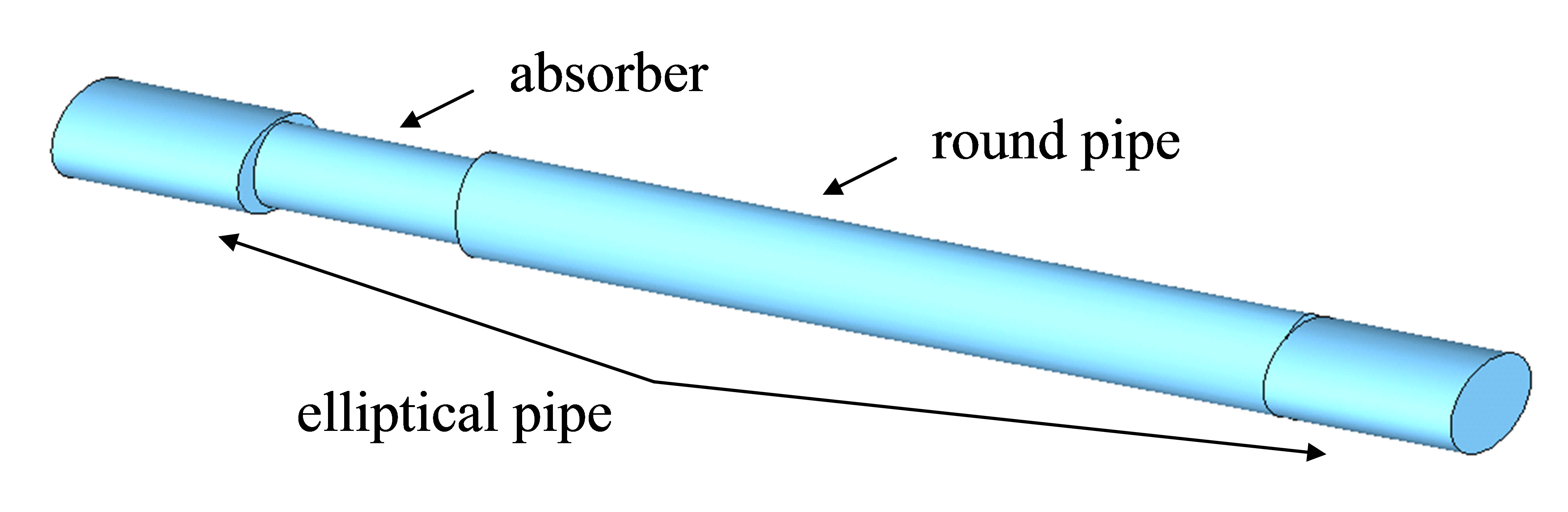}
	\caption{The geometry of the vacuum chamber in the undulator intersections.}
	\label{Fig02}
\end{figure}
Let us apply this method to the undulator intersection at the European XFEL. Here the vacuum chamber changes from an elliptical pipe to a round one. At the position of the elliptical-to-round transition (E2R) an elliptical absorber of  a smaller cross-section is placed as shown in Fig.~\ref{Fig02}.
\begin{figure}[!htb]
	\centering
	\includegraphics*[width=350pt]{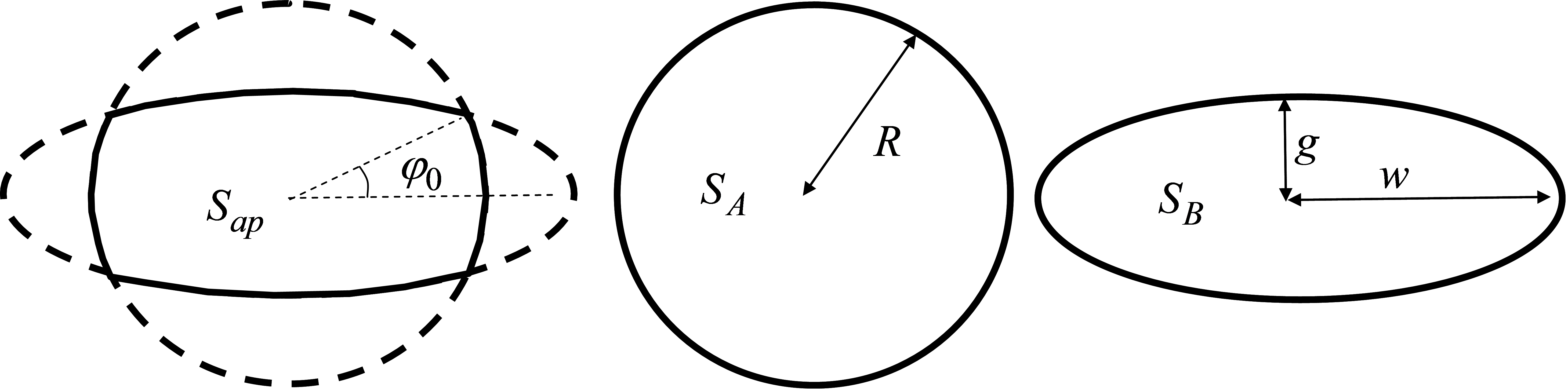}
	\caption{The geometry of transition from round pipe to elliptical one.}
	\label{Fig03}
\end{figure}
At the beginning let us consider a simple case without the absorber. In this case we have only an elliptical-to-round (E2R) pipe transition. The geometry of the transition is shown in Fig.~\ref{Fig03}. 

The Green’s function for the Laplacian inside the circle of radius $R$ can be written as
\begin{align}
\phi_R(\vec r_1,\vec r)=\frac{1}{4\pi\epsilon_0}\Re \left(2\ln\frac{z^*z_1-R^2}{R(z-z_1)} \right),\nonumber\\
\vec r=(x,y)^T,\quad,z=x+iy.\nonumber
\end{align}

The Green’s function for the Laplacian inside the ellipse with half axes $w$ and $g$ can be written as  ~\cite{Glucks}
\begin{align}
\phi_E(\vec r_1,\vec r)=\phi_E^0(\vec r_1,\vec r)-\phi_E^0(\vec r_1,\vec r_0),\quad \vec r_0=(0,g)^T,\nonumber\\
\phi_E^0(\vec r_1,\vec r)=-\frac{1}{\pi \epsilon_0} \left( \sum_{n=1}^{\infty}\frac{e^{-nu}}{n} F_n -\frac {1}{4}\ln ||\vec r -\vec r_1||^2\right),\nonumber\\
F_n=\frac{\Re T_n(\frac{x+iy}{d})\Re T_n(\frac{x_1+iy_1}{d})}{\cosh(nu)}+\frac{\Im T_n(\frac{x+iy}{d})\Im T_n(\frac{x_1+iy_1}{d})}{\sinh(nu)},\nonumber\\
d=\sqrt{w^2-g^2},\quad u=\coth^{-1}(w/g),\nonumber
\end{align}
where $T_n(z)$ are the Chebyshev polynomials of the first kind.

For the round-to-eliptical (R2E) pipe transition the Green’s functions has to be assigned as $\phi_A=\phi_R, \phi_B=\phi_E$. The aperture $S_{ap}$  is shown in Fig.~\ref{Fig03}.

The longitudinal impedance on the axis can be found as one-dimensional integral
\begin{align}\label{EqRE}
Z_{\parallel}^{R2E}=\frac{4}{c\pi}\int_{0}^{\varphi_0}\phi_E(\varphi,R)d\varphi,\\ \varphi_0=\tan^{-1}\left(\frac{g}{w}\sqrt{\frac{w^2-R^2}{R^2-g^2}}\right),\nonumber
\end{align}			
where $\phi_E(\varphi,r)\equiv \phi_E(\vec 0, (r\cos(\varphi),r\sin(\varphi))^T)$.

In order to calculate the longitudinal impedance of the elliptical-to-round (E2R)  pipe transition we can use the directional symmetry relation from~\cite{Zag06}
\begin{align}
Z_{\parallel}^{E2R}=Z_{\parallel}^{R2E}-\frac{2}{c}[\phi_E(\vec 0,\vec 0)-\phi_R(\vec 0,\vec 0)].\nonumber
\end{align}	
			 
\begin{figure}[!htb]
	\centering
	\includegraphics*[width=350pt]{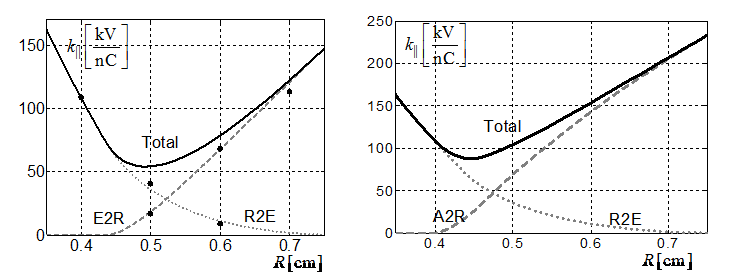}
	\caption{Dependence of the loss factor from the radius of the round pipe. The left graph presents the results without the absorber, the right graph presents the results with the absorber included.  
		The black dots show the numerical results from CST Particle Studio.}
	\label{Fig04}
\end{figure}

We evaluate the one dimensional integral, Eq.~\ref{EqRE}, numerically. The right graph in  Fig.~\ref{Fig04} presents the results for the fixed size of the elliptical pipe ($w=7.5\mathrm{mm}, g=4.4\mathrm{mm}$) and the Gaussian beam with rms length $\sigma_z=25 \mathrm{\mu m}$. The black dots show the numerical results from CST Particle Studio~\cite{CST} obtained for the bunch length  $\sigma_z=100\mathrm{\mu m}$ and scaled  to the bunch length $\sigma_z=25 \mathrm{\mu m}$.

Let us now consider the geometry with the absorber included. Here we consider the absorber as a long collimator. The absorber has the half width $w_1=4.5\mathrm{mm}$   and the half height $g_1=4\mathrm{mm}$. The transition from the elliptical pipe to the absorber (E2A) can be considered as in-step transition and we have $Z_{\parallel}^{E2A}=0$.
The contribution of the absorber to round pipe (A2R) transition can be found from Eq.~(\ref{EqOptImp}) with $w_1$  and $g_1$. The final result is presented in Fig.~\ref{Fig04} in the right graph. We can conclude that the optimal radius of the round pipe in the undulator intersection is 45-50 mm.

In the example considered the longitudinal impedance is written as one-dimensional integral Eq.~(\ref{EqRE}). The transverse impedance dipole and quadrupole terms in the Taylor expansion can be written in closed analytical form ~\cite{Opt07_2}. 

The application of the optical approximation to estimate the high frequency impedances of different transitions in the vacuum chamber of the European XFEL can be found in~\cite{Opt11}. The bunch used for the European XFEL operation is very short and the analytical results obtained  are quite accurate approximations to the coupling impedances. Most analytical results presented in~\cite{Opt11} are new and supplement those already published in~\cite{Opt07_2}.  The method of the optical approximation is powerful and allows to study analytically a truly large class of transitions when the analytical form of 2D Green functions of the pipe cross-sections are known. 

Usually the vacuum elements in the accelerators are connected with round, elliptical or rectangular pipes, for which the analytical Green functions are well known. For a general case the Green functions can be found through numerical solution of 2D Poisson's equations.

\section{Diffraction Model}

The optical theory ignores diffraction effects. It predicts zero impedance for the pillbox cavity or periodic array of irises. Indeed, in this case, all the three cross-sections $S_A$,
$S_\mathrm{ap}$ and $S_B$ are equal and Eq. (\ref{EqOptImp}) immediately gives a zero
result. 

\begin{figure}[!htb]
	\centering
	\includegraphics*[width=250pt]{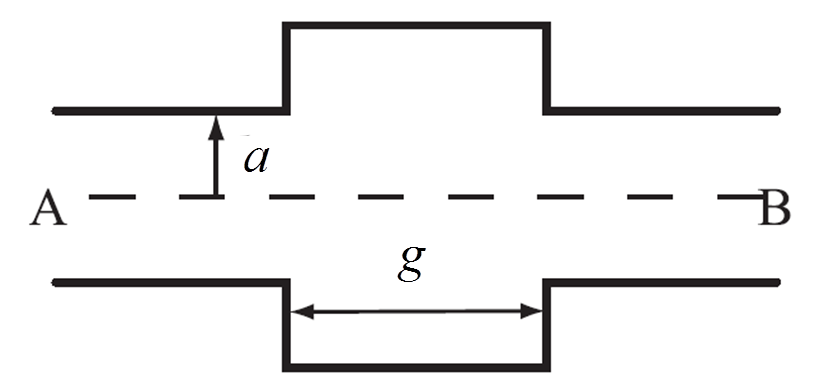}
	\caption{An axysimmetric deep pillbox cavity.}
	\label{Fig05}
\end{figure}

The diffraction theory takes into account the fact that radiated electromagnetic fields do not propogate along straight line. A Fresnel type integral from the diffraction theory of light is used to evaluate the electromagnetic energy that enters into the cavity region. This energy is associated with the energy lost by the beam and is thus related to the real part of the impedance.

For the deep pillbox cavity shown in Fig.~\ref{Fig05} the diffraction theory gives the high-frequency longitudinal (on the axis) impedance  
as (see, {\it e.g.} \cite{Chao93})
\begin{align}\label{diffraction_impedance}
Z_{\parallel}(k)
=
\frac{Z_0(1+i)}{2\pi^{3/2}a}
\sqrt{\frac{g}{k}},
\end{align}
where $a$ is the pipe radius and $g$ is the length of the cavity.

The reason for the optical approximation not reproducing the
result of the diffraction theory is that Eq.
(\ref{diffraction_impedance}) corresponds to the next order
approximation in the small parameter $\sigma_z g/a^2$~\cite{Opt07_1}.

For axysimmetric geometry the transverse impedance near the axis can be approximated as
\begin{equation}
\vec{Z}_{\perp}(k)=(Z_x,Z_y)^T=Z_d(k)(x_0,y_0)^T, \nonumber
\end{equation}
where $x_0,y_0$ are coordinates of source particle. The diffraction model at high-frequencies gives~\cite{Chao93}

\begin{align}\label{trans_diffraction_impedance}
Z_d(k)=\frac{Z_0(1+i)}{2\pi^{3/2}a^3k}\sqrt{\frac{g}{k}},
\end{align}
The corresponding wake  functions in the time domain read~\cite{Bane87}
\begin{align}
w_{\parallel}(s)=\frac{Z_0 c}{\sqrt{2}\pi^{2}a}\sqrt{\frac{g}{s}},\qquad\nonumber
w_d(s)=\frac{Z_0 c 2^{1.5}}{\pi^{2}a^3}\sqrt{g s}.\nonumber
\end{align}
For the Gaussian bunch with rms length $\sigma_z$ we can easily to calculate the loss and the kick factors
\begin{align}
k_{\parallel}=\int w_{\parallel}(s)\lambda(s)ds=\frac{Z_0 c}{4\pi^{2.5}a}\Gamma(0.25)\sqrt{\frac{g}{\sigma_z}},\nonumber\\
k_d=\int w_d(s)\lambda(s)ds=\frac{Z_0 c 2}{\pi^{2.5}a^3}\Gamma(0.75)\sqrt{g \sigma_z},\nonumber
\end{align}
where $\lambda(s)$ is the Gaussian charge density and $\Gamma$ is a gamma function.  

The same estimations for the impedances of pillbox cavity are obtained from parabolic equation method in~\cite{Stupakov2006}. 

\begin{figure}[!htb]
	\centering
	\includegraphics*[width=250pt]{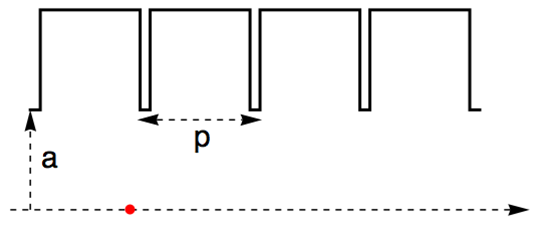}
	\caption{Periodic array of deep cavities.}
	\label{Fig06}
\end{figure}

The longitudinal impedance of one isolated pillbox cavity, Eq.(\ref{diffraction_impedance}), has $k^{-0.5}$ high frequency behavior. For an array of cavities with period $p$ (see Fig.\ref{Fig06}) the high frequency behaviour is quite different. It scales as $k^{-1.5}$. The high frequency impedance of an infinite cavity array was found in~\cite{Gluckstern1989, Bane1999} and it reads

\begin{align}\label{RoundImp}
Z_{\parallel}(k)=\frac{Z_0}{2\pi a}\left[\frac{1}{\eta(k)}-ik\frac{a}{2}\right]^{-1},\\
\eta(k)=\left[\frac{1-i}{2}\alpha\left(\frac{g}{p}\right)p\sqrt{\frac{k\pi}{g}}\right]^{-1},\label{EqBane}\\
\alpha(x)=1-0.465\sqrt{x}-0.070x.\nonumber
\end{align}

Inverse Fourier transforming, one obtains an analytical expression for the wake function:
\begin{align}
w_{\parallel}^{(1)}(s)=-\frac{Z_0 c}{\pi a^2} e^{s/s_0} \text{erfc}(\sqrt{s/s_0}),
\end{align}
with the distance scale factor $s_0=a^2g/(2\pi \alpha^2 p^2)$.

\begin{figure}[!htb]
	\centering
	\includegraphics*[width=250pt]{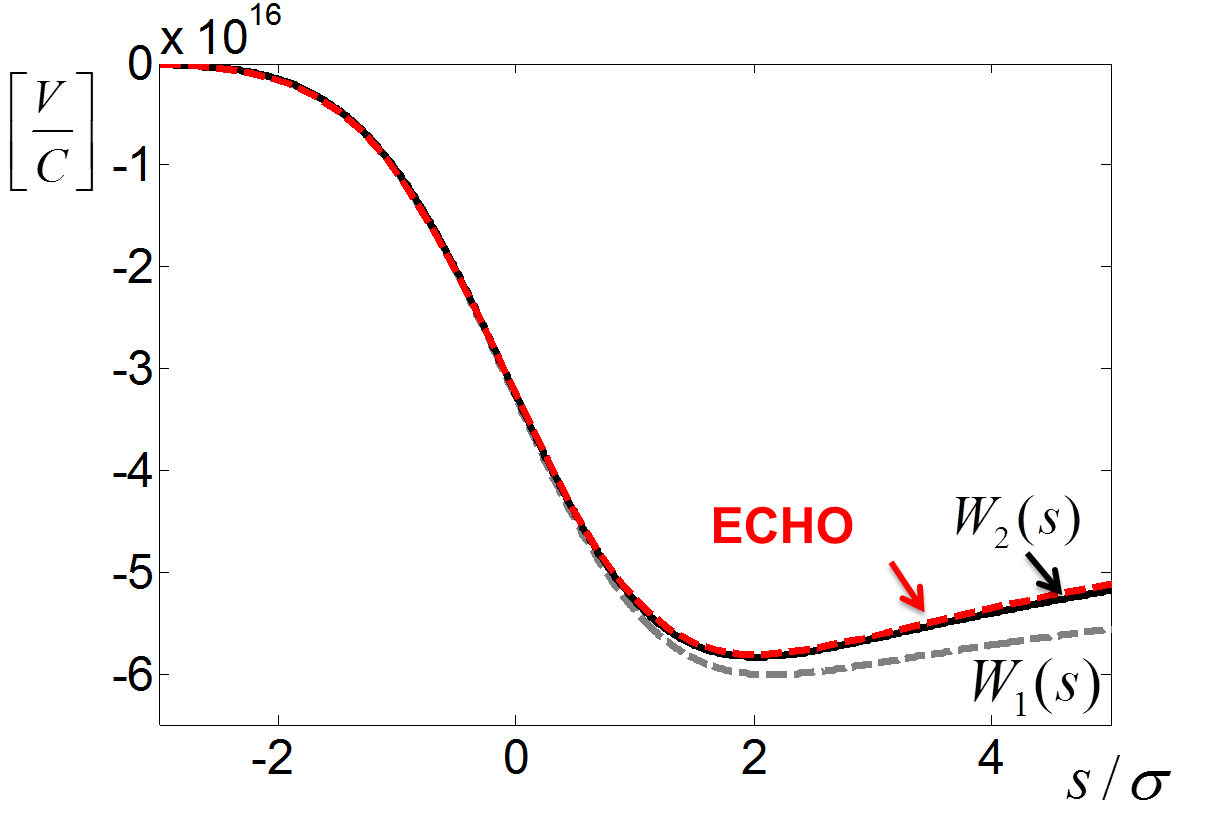}
	\caption{Longitudinal wake of periodic array of thin diaphragms (g/p=1).}
	\label{Fig07}
\end{figure}
\begin{figure}[!htb]
	\centering
	\includegraphics*[width=250pt]{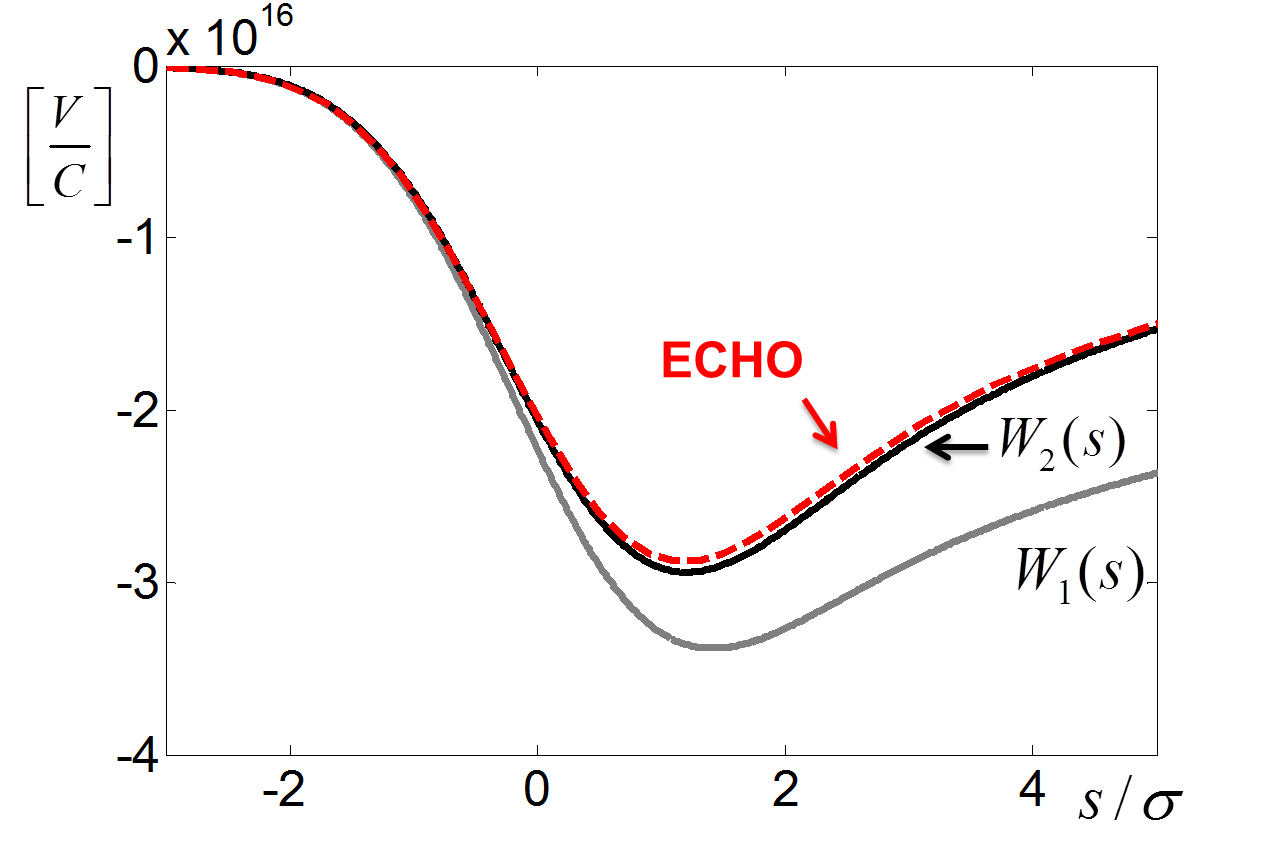}
	\caption{Longitudinal wake of  periodic array of short cavities (g/p=0.1).}
	\label{Fig08}
\end{figure}

We have compared this analytical estimation, Eq.(\ref{RoundImp}), with accurate numerical solution of Maxwell's equations by code ECHO~\cite{ECHO}. We will reference the numerical solution as "exact" one. 

We consider a chain of pillbox cavities with period $p=0.5\mathrm{mm}$. The cavities have radius $b=1.2\mathrm{mm}$ and are joined with a pipe of radius $a=0.7\mathrm{mm}$. The bunch is Gausian with rms length $\sigma_z=10\mathrm{\mu m}$. 

Our first example is a structure with periodic array of thin
diaphragms, $g/p=0.98$. Fig.~\ref{Fig07} shows the longitudinal wake potential
\begin{align}
W_{\parallel}(s)=\int_{-\infty}^{s}w_{\parallel}(s')\lambda(s-s')ds'.
\end{align}
The dashed curve labeled as "ECHO" is the numerical result, the gray dashed curve labeled as "$W_1(s)$" is the analytical result, Eq.(\ref{RoundImp}), which disagrees slightly with the "exact" solution.  Fig.~\ref{Fig08} shows the results for another case where the cavity gap is much smaller than the period, $g/p=0.1$. In this case the disagreement between the "exact" solution and the approximation, Eq.(\ref{RoundImp}), is large.

For the case of periodic array of infinitely thin diaphragms, $g/p=1$, an accurate approximation of the impedance was found earlier by G. Stupakov~\cite{Stupakov96}. In the high frequency approximation it gives
\begin{align}\label{EqStup}
	\eta(k)=\left[\frac{1-i}{2}\alpha(1)\sqrt{p k\pi}+\frac{1}{2}\right]^{-1}.
\end{align}
It can be seen that Stupakov's solution contains the additional term, which improves the agreement with the "exact" solution considerably. We would like to have the same order term in the more general case for arbitrary gap $g<p$. We combine Eq.(\ref{EqBane}) with Eq.(\ref{EqStup}) and suggest a more general equation
\begin{align}\label{EqZag}
\eta(k)=\left[\frac{1-i}{2}\alpha\left(\frac{g}{p}\right)p\sqrt{\frac{k\pi}{g}}+\frac{1}{2}\frac{p}{g}\right]^{-1}.
\end{align}
This equation differs from Eq.(\ref{EqBane}) by additional term $p/(2g)$ and it reduces to Stupakov's result for infinitely thin irises, $g=p$.

There is no exact Fourier transform of this impedance. We introduce here an approximate wake function:
\begin{align}
w_{\parallel}^{(2)}(s)=-\frac{Z_0 c}{\pi a^2} e^{-\sqrt{s/s_1}-s/s_2},\nonumber\\
s_1=s_0\frac{\pi}{4},\quad s_2=s_1\left(\frac{1}{2}-\frac{\pi}{4}+\frac{s_1 p}{ag}\right)^{-1}.\nonumber
\end{align}
This wake function has the same Taylor expansion up to the third order as the exact Fourier transform of Eq.(\ref{EqZag}). The corresponding wake potentials labeled as "$W_2$" are shown by black solid lines in Figs.~\ref{Fig07}-~\ref{Fig08}. It can be seen a good agreement with the "exact" numerical solution. 

The high frequency transverse impedance is related to the longitudinal impedance according to $Z_d=2Z_{\parallel}/(ka^2)$. Hence the transverse dipole wake function can be found as
\begin{align}
w_d^{(2)}(s) =\frac{2}{a^2}\int_{-\infty}^{s}w_{\parallel}^{(2)}(s')ds'.\nonumber
\end{align}

In the next section we will argue that the introduced function $\eta(k)$ can be treated as a surface impedance of corrugated waveguide of arbitrary  cross-section.

\section{Surface Impedance}

The impedance of a round metallic pipe of radius $a$ with conductivity $\kappa$ has long been known~\cite{Chao93} and is given by Eq.~(\ref{RoundImp}) with resistive surface impedance
\begin{align}
\eta=\eta^c=\frac{1}{Z_0}\sqrt{\frac{i\omega\mu}{\kappa}},\quad \omega=kc.\nonumber
\end{align}

Let us consider a case when the elements of the vacuum chamber that generate the beam impedance are small and uniformly distributed over the surface of the wall. One example of such an impedance is that due to surface roughness. Another example is a corrugated structure~\cite{Stupakov12}. While exact calculation of the impedance in such cases is difficult the effect on the beam can often be represented by a surface impedance. In the accelerator context the surface impedance was previously employed by Balbekov for the treatment of small obstacles in a vacuum chamber~\cite{Balbekov93}. For a rough surface it was introduced by Dohlus~\cite{Dohlus01}.

It was shown in~\cite{Dohlus01,Tsakanian09} that the effect of the oxide layer and the roughness can be taken into account through the inductive part of the surface impedance
\begin{align}
\eta=\eta^c+i\omega\frac{L}{Z_0},\nonumber\\
L=\mu_0((1-\epsilon_r^{-1})d_{oxid}+0.01d_{rough}),\nonumber
\end{align}
where $d_{oxid},\epsilon_r$ are the thickness of the oxid layer and it's relative permittivity, $d_{rough}$ is a rms roughness parameter~\cite{Dohlus01}.

If the surface impedance is known then we can consider an arbitrary (smooth enough) cross section of waveguide with the impedance boundary condition.

Let us consider a structure having rectangular cross section, where the material at top and bottom can vary as function of longitudinal coordinate but the width and side walls remain fixed and are perfectly electric conducting.
The impedance of such structure of halfwidth $w$ (in $x$-direction) can be written as~\cite{Zag15}
\begin{align}\label{EqRectImp}
Z_{\parallel}(k)=\frac{1}{w}\sum_{m=1}^{\infty}Z(y_0,y,k_x^m,k)\sin(k_x^m x_0)\sin(k_x^m x),\\ 
Z(y_0,y,k_x,k)=Z^c(k_x,k)\cosh(k_x y_0)\cosh(k_x y)+\nonumber\\ Z^s(k_x,k)\sinh(k_x y_0)\sinh(k_x y)\nonumber,
\end{align}
where $k_x^m=(\pi m)(2w)^{-1}$ is a transverse harmonic number. 

Using the surface impedace for calculating of high frequency impedance in flat geometry was considered in~\cite{Bane15, Bane16}. It was found that the coefficients in Eq.(\ref{EqRectImp}) can be written in the form
\begin{align}
Z^c(k_x,k)=\frac{Z_0c}{2a}\text{sech}^2(ak_x)\left[\eta^{-1}-ika\frac{\tanh(ak_x)}{ak_x}\right]^{-1},\nonumber\\
Z^s(k_x,k)=\frac{Z_0c}{2a}\text{csch}^2(ak_x)\left[\eta^{-1}-ika\frac{\coth(ak_x)}{ak_x}\right]^{-1}.\nonumber
\end{align}
If we use Eq.(\ref{EqBane}) or Eq.(\ref{EqZag}) for the surface impedance in the last expressions  then we obtain the high-frequency impedance of the rectangular corrugated structure. Following this approach analytical approximations for the wake functions in  flat/rectangular corrugated structures are derived in~\cite{Bane16,Bane16_2, Bane16_3}.

\section{Wake Asymptotics at the Origin}

The limit of high frequencies corresponds to small
distances behind a point charge. For infinitely long cylindrically symmetric disk-loaded accelerator structure, the steady-state wakes at the origin is
\begin{align}
w_{\parallel}(0^+)=-\frac{Z_0 c}{\pi a^2},\quad
\frac{\partial}{\partial s} w_{\perp}(0^+)=\frac{2 Z_0 c}{\pi a^4}.\nonumber
\end{align}

The same is true for a resistive pipe, a pipe with small periodic corrugations, and a dielectric tube within a pipe. It was assumed in~\cite{Bane03_1, Bane06} that it is generally true. For a non-round structure the constants are different, but again dependent only on transverse dimensions and independent of material properties. This statement for an arbitrary slow down layer was rigorously derived in~\cite{Bat14, Bat16} and it was shown there that for planar, square, and other cross section geometries, one can obtain a corresponding form factor coefficient by using a conformal mapping of these shapes onto the disk.

The asymptotics of the wakes at origin for a short transition and an isolated cavity are different. We summarize the asymptotic behavior at the origin of the considered models in Table~\ref{Table01}.
\begin{table}[htbp]
	\centering
	\caption{Asymptotics of wake functions at the origin}
	\label{Table01}
	\begin{tabular}{lcc}
		{\bf Model}&  ${\bf w_{\parallel}(s) }$ & $ {\bf w_{\perp}(s)}$\\
	    Optical (short transition)    & $\delta(s)$ & $O(1)$ \\
		Diffraction (cavity)     & $1/\sqrt{s}$ & $\sqrt{s}$ \\
		Diffraction (cavity chain)   & $O(1)$ & $O(s)$ \\
		Slow down layer   & $O(1)$ &$O(s)$ \\
	\end{tabular}
\end{table}

\section{Combining Computations and Analytics}

\begin{figure}[!htb]
	\centering
	\includegraphics*[width=350pt]{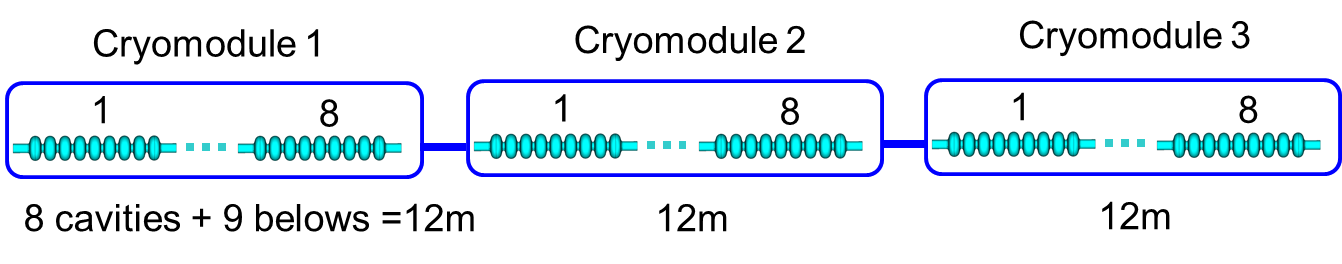}
	\caption{Three TESLA cryomodules.}
	\label{Fig09}
\end{figure}

The real geometry of accelerator vacuum chamber is quite complicated and it is a challenge to give a short range wake functions for it. However in many situations it is possible to combine the considered above analytical models with numerical computations. 

The first possibility is to take an analytical model for a simple geometry and to assume that the real vacuum chamber can be described by the same model with different coefficients. These coefficients can be found from fitting of the model to results of numerical simulations. Such approach was elaborated in~\cite{Novo99, Zag03} in order to estimate the wake functions in TESLA linac of the Eropean XFEL and FLASH at DESY. 

\begin{figure}[!htb]
	\centering
	\includegraphics*[width=350pt]{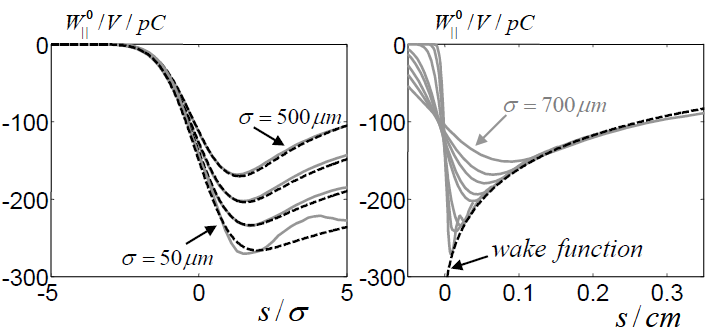}
	\caption{Comparison of analytical and numerical longitudinal wake potentials in the third cryomodule.}
	\label{Fig10}
\end{figure}
\begin{figure}[!htb]
	\centering
	\includegraphics*[width=350pt]{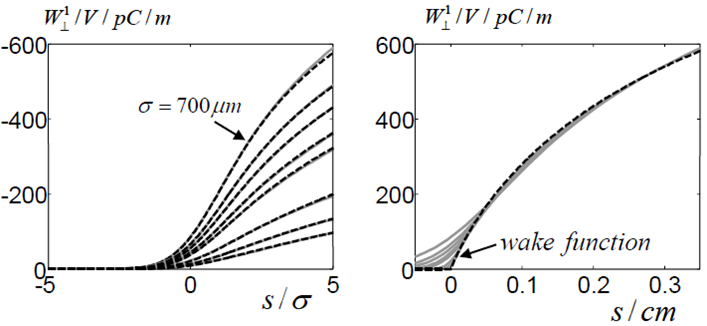}
	\caption{Comparison of analytical and numerical transverse wake potentials in the third cryomodule.}
	\label{Fig11}
\end{figure}

The TESLA linac consists of a long chain of cryomodules. The cryomodule of total length 12m contains 8 cavities and 9 bellows as shown in Fig.~\ref{Fig09}. The iris radius is 35mm and beam tubes radius is 39mm. The wakefields for Gaussian bunches up to $\sigma_z = 50\mathrm{\mu m}$ have been studied. In order to reach the steady state solution the structure of 3 cryomodules with total length 36m was considered. It was shown that as for periodic structure the loss factor becomes independent from the bunch length and the kick factor decreases linear with the bunch length (see Table~\ref{Table01}).

After fitting of coefficients in Bane's model~\cite{Bane03} the following wake functions (for one cryomodule) are obtained
\begin{align}
w_{\parallel}(s)=-344e^{-\sqrt{s/s_0}}\left[\frac{V}{pC}\right],\quad s_0=1.74\mathrm{mm}, s_1=0.92\mathrm{mm},\nonumber\\
w_{\perp}(s)=10^3\left(1-\left(1+\sqrt{\frac{s}{s_1}}\right)e^{-\sqrt{\frac{s}{s_1}}}\right)\left[\frac{V}{pC m}\right].\nonumber
\end{align}
Fig.~\ref{Fig10} shows numerical (gray solid lines) and analytical (black dashed lines) wake potentials for bunches with $\sigma_z = 500,250,125,50\mathrm{\mu m}$. The deviation of the curves for the shortest bunch can be explained by insufficiency of the 3 cryomodules to reach the steady state solution. At the right side of Fig.~\ref{Fig10}  the wakes (gray solid lines) together with the analytical wake function (black dashed line) are shown. The analytical wake function tends to be the envelope function to all wakes. Fig.~\ref{Fig11}  shows likewise the results for transverse wakes.

It can be seen from Table~\ref{Table01} that the behavior of wake functions for infinite periodic structure and for isolated cavity are different. In~\cite{Zag04} we have combined two models in order to obtain wake functions of high harmonic module and transverse deflecting structure used at FLASH facility at DESY.

Recently another method was suggested in~\cite{Pod13}. The idea behind the method is to use a combination of computer simulations with an analytical form of the wake function for a given geometry in the high-frequency limit (optical or diffraction model). For example, the longitudinal wake function of round step-out transition can be well aproximated as
\begin{align}
w_{\parallel}(s)=w_{opt}(s)+d(s),\nonumber\\
w_{opt}(s)=-\frac{1}{\pi\epsilon_0}\ln(ba^{-1})\delta(s),\nonumber\\
d(s)=(\alpha+\beta s).\nonumber
\end{align}
The crucial element of the method is that the smooth function  $d(s)$ can be obtained from simulations with long bunch by fitting to the formula. This method that combines a (processed) long-bunch wake from an EM solver and a singular analytical wake model allows one to accurately obtain wake fields of short bunches, including that of a point-charge.

\section{Impedance Database Model and Beam Dynamics Simulations}

The European XFEL contains hundreds of sources of the coupled impedances. In order to obtain the wake functions of different   elements   we   have used   analytical   and  numerical   methods.  The  wake  functions of relativistic charge  have  usually  singularities  and  can  be    described    only    in terms    of    distributions  (generalized  functions).  An approach to tabulate such functions and use them later to obtain wake potentials for different bunch shapes was introduced in~\cite{ZagDB,Dohlus12,Zag16}.

The longitudinal wake function near the reference trajectory $\vec{r_a}$ can be presented through the second order Taylor expansion
\begin{align}
w_z(\vec{r},s)=w_z(\vec{r_a},s)+<\nabla w_z(\vec{r_a},s),\Delta \vec{r}>+\nonumber\\
 \frac{1}{2}<\nabla^2 w_z(\vec{r_a},s) \Delta \vec{r},\Delta \vec{r}>+O(\Delta \vec{r}^3),\nonumber
\end{align}
where we have incorporated in one vector the transverse coordinates of the source and the witness particles, $\vec{r}=(x_0,y_0,x,y)^T$, $\Delta \vec{r}=\vec{r}-\vec{r_a}$, and $s$ is a distance between these particles. 

For arbitrary geometry without any symmetry the Hessian matrix $\nabla^2 w_z(\vec{r_a},s)$ contains 8 different elements:
\begin{align}\label{Eq13}
\nabla^2 w_z(\vec{r_a},s)&=
\begin{pmatrix}
h_{11}&h_{12}&h_{13}&h_{14}\\
h_{12}&-h_{11}&h_{23}&h_{24}\\
h_{13}&h_{23}&h_{33}&h_{34}\\
h_{14}&h_{24}&h_{34}&-h_{33}
\end{pmatrix},\nonumber
\end{align}
where we have taken into account the harmonicy of the wake function in coordinates of the source and the witness particles~\cite{Zag15}. 

Hence in general case we use 13 one-dimensional functions to represent the longitudinal component of the wake function for arbitrary offsets of the source and the wittness particles near to the reference axis. For each of this coefficients we use the representation~\cite{ZagDB}
\begin{equation}\label{Eq14}
h(s)=w_0(s)+\frac{1}{C}+R c \delta(s)+c\frac{\partial}{\partial s} \left(L c \delta(s)+w_1(s)\right),
\end{equation} 
where $w_0, w_1 $ are non-singular functions, which can be tabulated easily and  constants $R, L, C$ have  meaning  of  resistivity,  inductance and capacitance, correspondingly. 
    The  wake  potential  for  arbitrary  bunch  shape  
$\lambda(s)$ can  be found by formula 
\begin{align}\label{Eq14}
W_h(s)=w_0*\lambda(s)+\frac{1}{C}\int_{-\infty}^{s}\lambda(s')ds'+R c \lambda(s) +\nonumber\\
c^2 L \lambda'(s)+c w_1(s)*\lambda'(s),\nonumber
\end{align} 
where $\lambda'$ is a derivative of $\lambda'$.

In order to model the beam dynamics in the presence of wakefields we use the open source code OCELOT \cite{Agapov2015}. We have developed and tested the wakefield module. The implementation follows closely the approach described in~\cite{ZagDB},~\cite{Dohlus12}.
The wakefield impact on the beam is included as series of kicks. In ~\cite{Zag16} we have studied a possibility to extend the bandwidth of the radiation at the European XFEL with the help of a special compression scenario together with the corrugated structure insertion. We have derived an accurate modal representation of the wake function of corrugated structure  and have applied this fully three dimensional wake function in beam dynamics studies with OCELOT in order to estimate the change of the electron beam properties.

\end{document}